\documentclass[]{spie}  

\usepackage{amsmath,amsfonts,amssymb}
\usepackage{graphicx}
\usepackage[colorlinks=true, allcolors=blue]{hyperref}
\usepackage{float}
\usepackage{caption}
\usepackage{subcaption}

\title{NAIR-APREXIS: Enabling photonics-based instruments for long-baseline interferometry and integral-field spectroscopy}

\authorinfo{Further author information: Lucas Labadie: E-mail: labadie@ph1.uni-koeln.de}

\author[a]{L.~Labadie}
\author[b]{R.~J.~Harris}
\author[c]{K.~Madhav}
\author[c,f]{A.~Dinkelaker}
\author[a]{K.~Barjot}
\author[d]{A.~Beno\^it}
\author[e]{N.~J.~Scott}
\author[e]{N.~Anugu}
\author[b]{S.~Mahdizadeh}
\author[b]{V.~Kutnohorsky}
\author[b]{A.~Calcines Rosario}
\author[b]{E.~Ronson}
\author[b]{A.~Magniez}
\author[d]{R.~R.~Thomson}
\author[a]{T.~K.~Sharma}
\author[c]{A.~V.~Mayer}
\author[e]{G.~Schaefer}

\affil[a]{I. Physikalisches Institut, Universität zu Köln, Zülpicher Str. 77, 50937 Köln, Germany}
\affil[b]{Centre for Advanced Instrumentation, Department of Physics, Durham University, South Road, Durham DH1 3LE, UK}
\affil[c]{Leibniz-Institut f\"ur Astrophysik Potsdam, An der Sternwarte 16, D-14482 Potsdam, Germany}
\affil[d]{Institute of Photonics and Quantum Sciences, Heriot-Watt Univ., Edinburgh EH14\,4AS, UK}
\affil[e]{The CHARA Array of Georgia State University, Mount Wilson Observatory, Mount Wilson, CA 91023, USA}
\affil[f]{Corning Optical Communications GmbH \& Co. KG, Walther-Nernst-Str. 5, 12489 Berlin, Germany}

\begin{document} 
\maketitle

\begin{abstract}
    The NAIR project -- Novel Astronomical Instrumentation based on photonic light Reformatting -- aims at advancing photonic technologies for infrared long-baseline interferometry and precision spectroscopy. The rapid development of astrophotonics over the past decade has opened new pathways for astronomical instrumentation with unprecedented capabilities. We present results from NAIR that demonstrate the potential of the ultrafast-laser inscription (ULI) technique for fabricating remapping devices for a range of applications. We developed a single-mode integrated-optics astronomical K-band beam combiner, which we successfully tested on-sky, although using only one single baseline of the CHARA Array. Across several observing campaigns, the prototype exhibited excellent stability, achieving 1\% precision on the interferometric visibilities and a total on-sky throughput $>$40\%, with an achieved limiting magnitude of K$\approx$5 using the 1-m meter telescopes of CHARA and without external fringe tracking. We are also developing an integral field unit (IFU) designed for exoplanet detection and characterisation. This is due to be tested with MagAO-X in Chile in 2027. The IFU is based upon astrophotonic fiber technologies - two-photon-polymerized (TPP) lenslets, a custom multi-core fiber, and a ULI reformatter. We discuss our efforts to achieve contrasts of 10$^{-3}$ between adjacent spaxels whilst retaining throughput of $>$50\%.
    Finally, we discuss the work we are doing developing the next generation of astrophotonic technologies, including TPP micro-dispersers designed for low resolving power, high transmission applications. We achieve R$\sim$30 in a sub-mm package, showing viability for future use. These results emphasize the versatility and simplicity of integrated photonic approaches as a major advance in optical technologies for astronomical instrumentation.
\end{abstract}


\section{INTRODUCTION}\label{sec:introduction}

\begin{figure}
    \centering
    \includegraphics[width=0.80\linewidth]{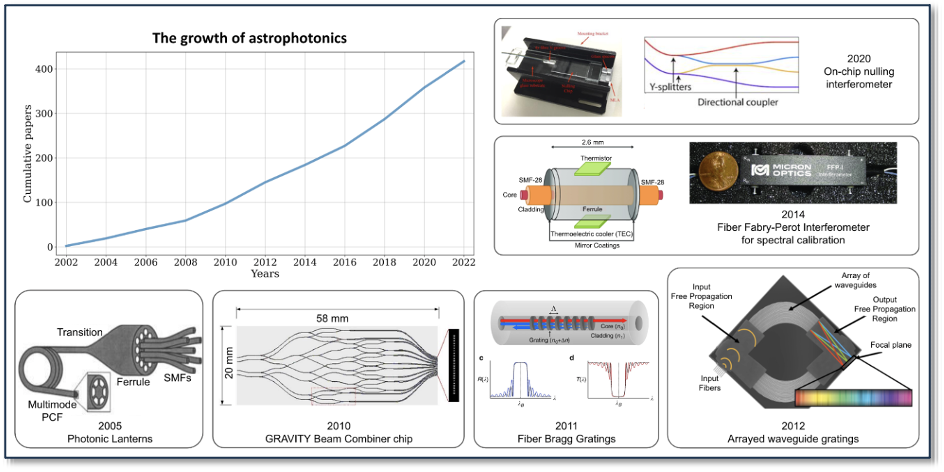}
    \caption{Observational techniques covered by astrophotonics and including long-baseline interferometry as well as high-resolution spectroscopy. The increase of the astrophotonics literature is also shown \cite{Jovanovic2023}\,.}
    \label{fig:Fig1}
\end{figure}

    In the last decade, the potential of photonics for astronomical applications has been increasingly demonstrated. 
Taking advantage of the well-established fields of telecommunications or THz photonics, the growth of astrophotonics affects many key functionalities of astronomical instrumentation (cf. Figure~\ref{fig:Fig1}). The main motivation for implementing photonic integrated circuits (PICs) in astronomy is their ability to control and manipulate light with higher versatility than possibly with classical bulk optics. A typical advantage of PICs is the strong reduction in the footprint size and the weight of the core functionalities, hence gaining significant stability. The cost of production of PICs can also, depending on the manufacturing platform used, be reduced. 
    The single-component, monolithic nature of PIC functionalities enables a higher temporal stability of the instrumental transfer function. This has been particularly visible in the second-generation instrument of the VLTI such as GRAVITY \cite{Eisenhauer2017}, whose interferometric unit is based on a silica-on-silicon integrated optics beam combiner \cite{Perraut2018}\,.

    One of the main advantages of PICs in astronomical instrumentation is in the incremental extension and upgrade of existing instruments it offers. For instance, a beam combiner or fiber-based remapper can be relatively easily exchanged after a new fabrication run at low-cost. This means that the turnaround between the development of a first-generation PIC and a first upgrade can be fast. 
    An example of the new dimension of astrophotonics lies in the effort that is put in this discipline in the context of the future Habitable World Observatory (HWO) with the AstroPIC project \cite{Sirbu2025}\,. 
    
    While astrophotonics shows a lot of promise in the advance of new concepts of astronomical instruments, we face challenges that are very specific to the requirements of astronomy. These are, for instance, the efficient interfacing with the imperfect, time-variable, Point Spread Function (PSF) of the telescope, as well as the wide spectral bandwidth corresponding to the different astronomical filters, which is typically limited by the strong chromatic nature of the waveguides. 
    In this context, the NAIR-APREXIS project focuses on three main areas of work covering a) long-baseline interferometry; b) integral field spectroscopy; c) integrated devices for broader applications in both of the two previous areas. 
    
    a) For long-baseline interferometry relying on coherent beam combination, the control of the phase and of the polarization is a critical aspect. The single-mode nature of the PIC devices employed ensures wavefront filtering that efficiently suppresses the wavefront phase aberrations that would otherwise hinder the achievable interferometric contrast. 
    While photonics play an important role for the remapping functions that are necessary to techniques such as aperture masking, the robust calibration of the transfer function is a delicate but necessary step for doing science, which requires an excellent understanding of the chromatic and polarization behavior of the underlying PIC when trying to cophase the beams.

    b) In the area of high-resolution integral field spectroscopy considered in NAIR-APREXIS, the objective is primarily to deliver efficient incoherent remapping devices that allow us to deliver a pseudo-slit feeding a spectrograph. Although in conventional instruments a slicer is employed, this element is optically complex and contributes to overall losses in the system. Having a slicer ``integrated" in a single fiber link is therefore an advantage when. Furthermore, the modal noise problem known in classical multi-mode fiber-fed spectrographs can be removed if working in single-mode/ diffraction-limited conditions.
    
    c) Dispersive elements in astrophotonic-based instrumentation are typically based on classical prism or gratings, depending on the desired spectral resolution. Significant progress in the level of integration can be obtained if the dispersive element can be mechanically attached to the PIC. This is investigated in NAIR-APREXIS as well.
    
    Although long-baseline interferometry and integral field spectroscopy cover in the first place a different parameter space, the NAIR-APREXIS project promotes synergies and cross-fertilization between both areas.
           
\section{Structure of the project}\label{sec:structure}
From the methodology point of view, NAIR-APREXIS considers three major parts.

\subsection{Fabrication platform}

\begin{figure}[H]
    \centering
    \includegraphics[width=0.60\linewidth]{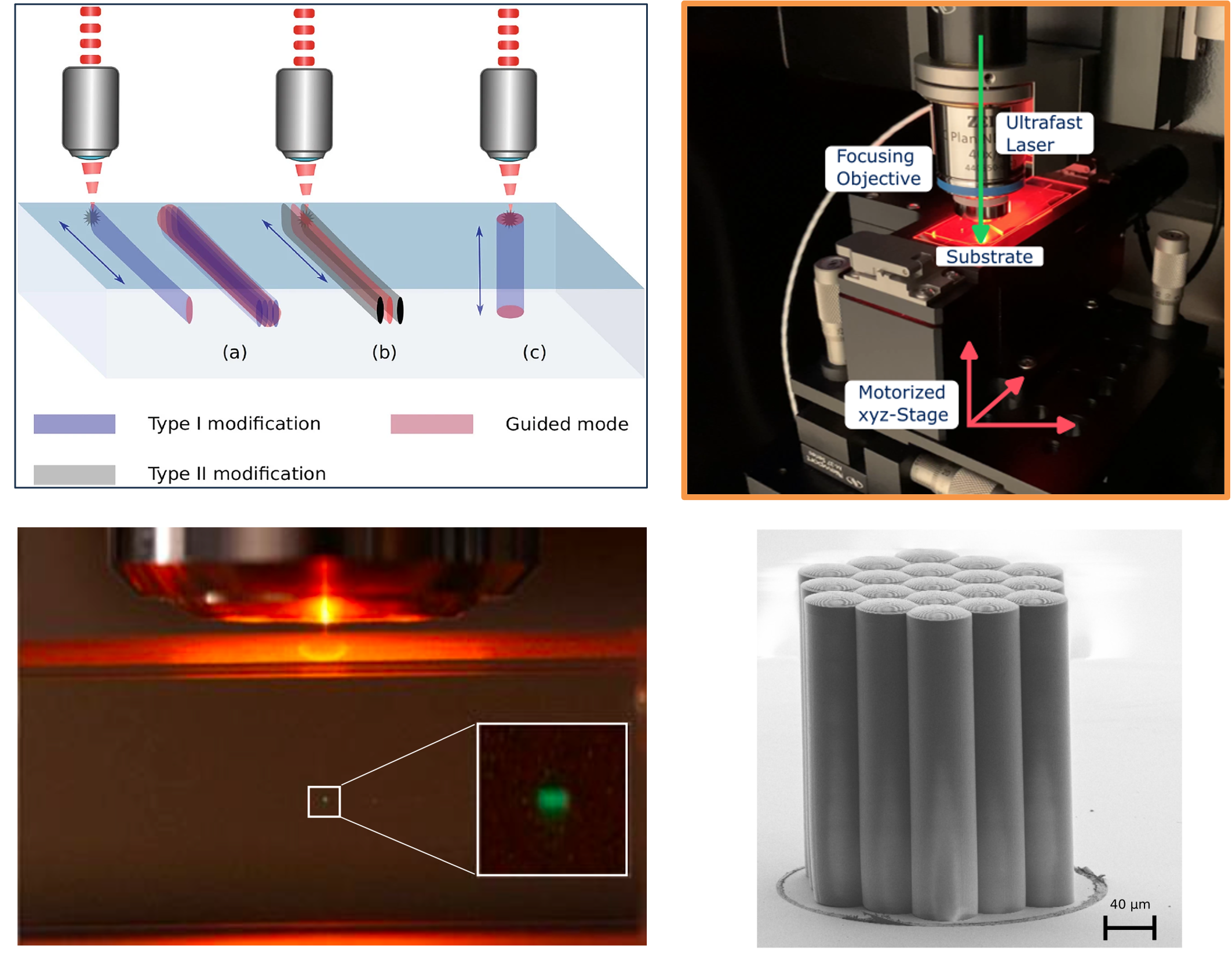}
    \vspace{0.1cm}
    \caption{Top-left: functional principle of the technique of ultrafast laser writing involving Type\,I or Type\,II modification of the substrate \cite{Ghafur2023}\,. Bottom-left: Principle of two-photon absorption and polymerization that occur only at the focal region, enabling highly confined polymerization in 3D (Image credit: S. Ruzin and H. Aaron, UC Berkley). Top-right: example of ULI platform at AIP-Potsdam (Image credit: A. V. Mayer and A. Dinkelaker). Bottom-right: result of a 3D-printing run showing a micrometer-size scale lenslet array \cite{Anagnos2021}\,.}
    \label{fig:Fig2}
\end{figure}

The realization of astrophotonics instrument with good performance is based on the mastering of a technological platform adapted to the identified astronomical requirements. More information on manufacturing in the context of astrophotonics can be found in Labadie et al. 2016 \cite{Labadie2016} an reference therein. For instance, the GRAVITY integrated optics beam combiner has taken advantage of the high maturity of the silica-on-silicon platform \cite{Mottier1997}\,. In the recent past of the NAIR project, the use of the Ultrafast Laser Inscription platform has demonstrated to offer significant advantages, combining versatility in the chosen two- or three-dimensional waveguide networks together with the relative ease of realization (see Figure~\ref{fig:Fig2}, reviews \cite{Choudhury2014,Gross2015} and references therein). In NAIR, we try to maximize the potential return for astrophotonics of this fabrication platform. 

In essence, the ULI platform focuses through a well-chosen microscope objective an ultrafast laser beam into a glass substrate, triggering non-linear photon/matter interaction properties that locally modify the refractive index of the glass, hence generating a field-confining waveguide structure. Compared with the photolithography approach, the lower level of repeatability of this platform may require the production of batches containing several PICs to be tested and selected. However, significant progress have been made: ULI is now used in a large variety of glasses and waveguides can be written with identical lengths down to $\sim$10\,nm.

A variant of the ULI approach, though based on comparable physical principles, is the two-photon polymerization that allows to grow three-dimnensional printed nano- and micro-structures for light confinement, routing or dispersion with a precision compatible with the high level of integration required in astrophotonics. 

\subsection{Interface with the observational infrastructure}

The primary objective of NAIR is to enable instrumentation for astronomical science observations. Therefore, it is crucial that the collection of photons from the telescope is ensured with high efficiency. This is a well-known difficulty to be faced since we are generally in photon-starving conditions (hence, making the PIC throughput requirement a very stringent one), polychromatic conditions (hence, dealing with chromatic dispersion and wavelength-dependent modal behavior), as well as in conditions where we face a turbulent atmosphere for ground-based observations (hence, making the coupling of the telescope PSF into small-core waveguides a challenge). Different solutions are explored in NAIR. These encompass mode-control devices such as photonic lanterns and structured multicore fibers. The three-dimensional ULI-based pupil- or image-plane remappers are essential to convert the two-dimensional light distribution delivered by a telescope into a simpler one-dimensional arrangement that can benefit the dispersive element (typically the spectrograph) but also the configuration used with a camera to record the output. For interferometric beam combiners, single-mode fibers arranged into a glued V-groove remain the preferred solution for coupling and filtering the AO-corrected PSF of the telescope.

\subsection{On-sky demonstration and qualification}

Although NAIR-APREXIS focuses on photonic technologies, the project does not limit itself to laboratory demonstration of proof-of-concepts. The objective is to push the development chain up to an on-sky instrument or upgrade that allows us to test the performances for science grade observations. This is the reason why in NAIR the collaboration with partners having access to astronomical facilities is so important. We really want to demonstrate the added value of our photonics for cutting-edge astronomical observations. Examples of such partners are members of the interferometric CHARA Array in California or of the LBT in Arizona. 

\section{Results on photonic beam combiners for interferometry}

In the first phase of the NAIR project, we have focused on the development of a four-telescope beam combiner in the context of an aperture-masking/pupil remapping experiment \cite{Perrin2006}\,. The principle is to interfere coherently four sub-pupils of the primary mirror of a telescope selected in a non-redundant manner, similar to what is achieved with long-baseline interferometry and for which it can be shown that the retrieved angular resolution is $\sim$ $\lambda$/$2D$. To mitigate the throughput deficit of classical aperture masking experiments (which only capture the light corresponding to about 10\% of the primary mirror surface), photonic pupil remappers associated with beam combiners allow complete encoding of the whole entrance pupil corresponding to the primary mirror. This removes the throughput penalty in the retrieval process of the coherence function of the observed object. On the technology side, we concentrated our effort 
on ULI in borosilicate glasses, which was a well-mastered development by our Italian partner from CNR and Politecnico di Milano. The consequence was that the photonic device was developed for the H-band, at 1.65\,$\mu$m, i.e., a spectral range in which the majority of astrophotonic developments concentrate \cite{Labadie2022}\,. Based on the Discrete Beam Combiner (DBC) concept \cite{Minardi2012}\,, a four-telescope H-band prototype was demonstrated at the William Herschel Telescope where the coherence function of the star Vega was obtained \cite{Nayak2021}\,. 

 \subsection{The ULI two-telescope K band beam combiner}

In the current phase of the NAIR project, we explored the potential of ULI to deliver new chips operating outside the typical telecom spectral window of silica. Numerous astronomical science cases such as disk-bearing young stellar objects or young low-mass companions can be better addressed at longer wavelengths owing the cooler temperature of the emitting body (circumstellar dust in disks or very low-mass objects), which justifies the extension of our photonic PICs beyond $\sim$2\,$\mu$m. ULI in chalcogenide glasses has already demonstrated the viability of this platform up to 4\,$\mu$m with the NOTT project \cite{Sanny2026b}\,. 
The challenge here was to demonstrate our ability to develop a ULI beam combiner using Infrasil glass, which had been poorly explored in the past. Infrasil made from fused quartz has mechanical and thermal properties very similar to UV fused silica. The low OH content makes it ideal for applications in the near-infrared thanks to its transmission range spanning from the ultraviolet up to $\sim$3.5\,$\mu$m. Infrasil is also commercially available at a competitive price. 
The manufacturing of several beam combiners was undertaken at Heriot Watt University, aiming for high performance and customized splitting ratio between the different interferometric and photometric channels. Another challenge was set by the requirement of a flat spectral response of the combiner to enable future spectrally dispersed operation with a comparable flux per spectral channel. 

\begin{figure}[t]
    \centering
    \includegraphics[width=1\linewidth]{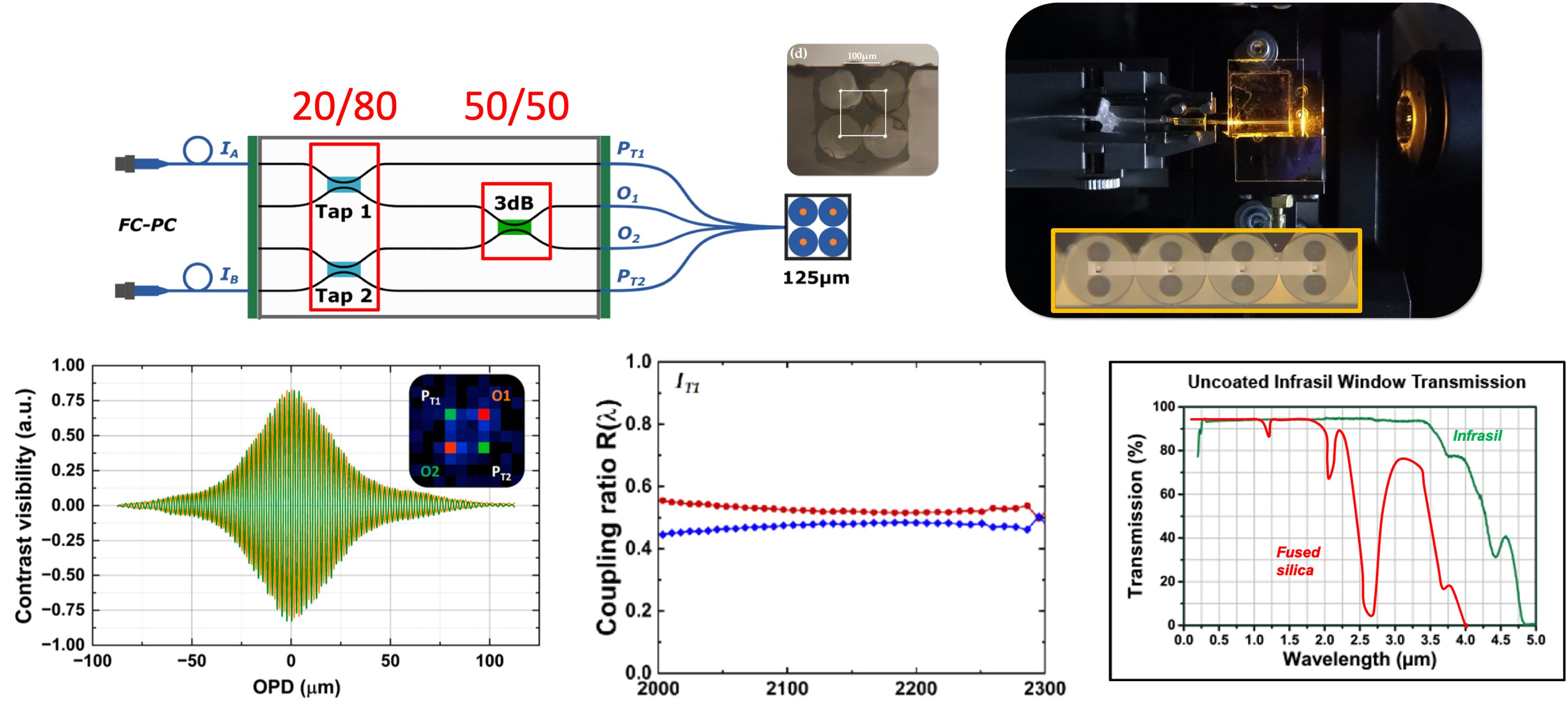}
    \vspace{0.25cm}
    \caption{2-telescope ULI K band beam combiner and its properties. In the top part, the beam combiner design is illustrated. It is made of cascaded directional couplers with different splitting ratios. The four outputs are assembled in a 2$\times$2 square arrangement, whereas the inputs are fed by a classical V-groove supporting four polarization maintaining fibers Nufern PM1950 with a conventional panda shape (right-hand side). The extracted interferograms are shown on the bottom part, together with the falt response of the coupling ratio \cite{Benoit2021,Siliprandi2024}\,. 
    The increased wavelength coverage of Infrasil glass compared to classical UV fused silica is highlighted on the right.}
    \label{fig:Fig3}
\end{figure}

Focusing on the astronomical K band between 2 and 2.5\,$\mu$m, we developed a two-telescope beam combiner in Infrasil glass, with the inputs connected to single-mode polarization-maintaining fibers. The photometric taps were designed to capture $\sim$20\% of the incident flux to enable simultaneous photometric calibration of the interferograms. The four outputs -- i.e., two interferometric channels and two photometric taps -- are routed with four fibers into a square 2$\times$2 arrangement (see below). As reported in different references \cite{Benoit2021,Siliprandi2024}\,, the laboratory characterization has demonstrated a nearly flat spectral response thanks to an asymmetric design of the directional couplers, together with an excellent throughput of about 80\% and an instrumental contrast above 82\% for the bare combiner (cf. Figure~\ref{fig:Fig3}). 
Although earlier approaches based on photolithography have demonstrated excellent performances such as in the case of the GRAVITY/VLTI beam combiner, we could demonstrate an alternative venue for the low-cost realization of K-band beam combiners. In the specific case of our two-telescope PIC, the advantage of 3D writing across the whole substrate volume \cite{Rodenas2012} was not exploited but it remains an important advantage when considering a larger number of telescope inputs. 

\subsection{On-sky tests at CHARA}

\begin{figure}[t]
    \centering
    \includegraphics[width=1\linewidth]{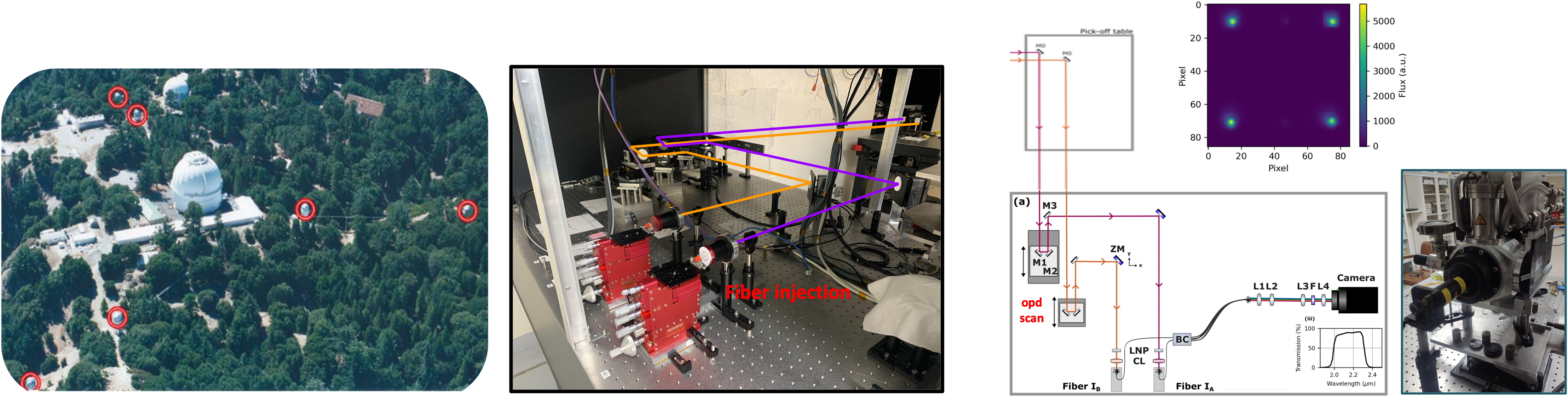}
    \vspace{0.25cm}
    \caption{Left: aerial view of the Mount Wilson Observatory in California. The large dome visible in the center is the one of the 100-inch Hooker telescope. Encircled in red are shown the fixed 1-m size telescope of the CHARA interferometric array. Center: close view of the CHARIOT table fed by two telescope beams shown in yellow and pink. Right: schematic view of the setup showing the pick-off table, the CRED-One camera and the four outputs of the beam combiner packed into a 2$\times$2 square arrangement.}
    \label{fig:Fig4}
\end{figure}

As elaborated in Section~\ref{sec:structure}, testing new technologies on-sky to prepare the definition of new astronomical instruments is a major goal of NAIR. For this purpose we have elaborated a collaboration with the CHARA Array, which currently hosts an interferometer offering the longest baselines inn the world for optical/infrared interferometry. The CHARA array hosts six 1-m fixed telescopes with baselines as long as 330\,m and one movable telescope with benefits from a fiber link to route the light towards the interferometric lab. An aerial view of the array is shown in Figure~\ref{fig:Fig4}. The ULI beam combiner is integrated within a new visitor instrument named CHARIOT, which offers a flexible platform to host photonic based instruments. CHARIOT takes advantage of the heritage of the FLUOR instrument \cite{Foresto1998} and, using single-mode photonics, concentrates on high-precision measurement of visibilities with errors dhown to $\sim$1\%.   

For this purpose, the optical table previously hosting FLUOR has been entirely refurbished, with the integration of new fiber couplers and a new high-sensitivity CRED-One camera for the H and K bands (for more details, see \cite{Mayer2024,Scott2024}). CHARIOT currently host two channels that can be upgraded to four in the future. The temporal encoding of the fringes is the current operational strategy, which also results from the FLUOR heritage. The fringes are scanned over $\sim$100\,$\mu$m thanks to a rapid delay line, which is complemented by a static delay for the search of the zero optical path delay (cf. the proceeding by K. Barjot in the conference 14148). Because of the pre-existing readout software of the FLUOR experiment, a square arrangement of the outputs of the chip had to be maintained. This was possible by taking advantage of the output V-groove delivering fibers that could be selectively arranged in the desired geometry, hence demonstrating the versatility and flexibility of the photonic approach. 

 \begin{figure}[b]
    \centering
    \includegraphics[width=0.85\linewidth]{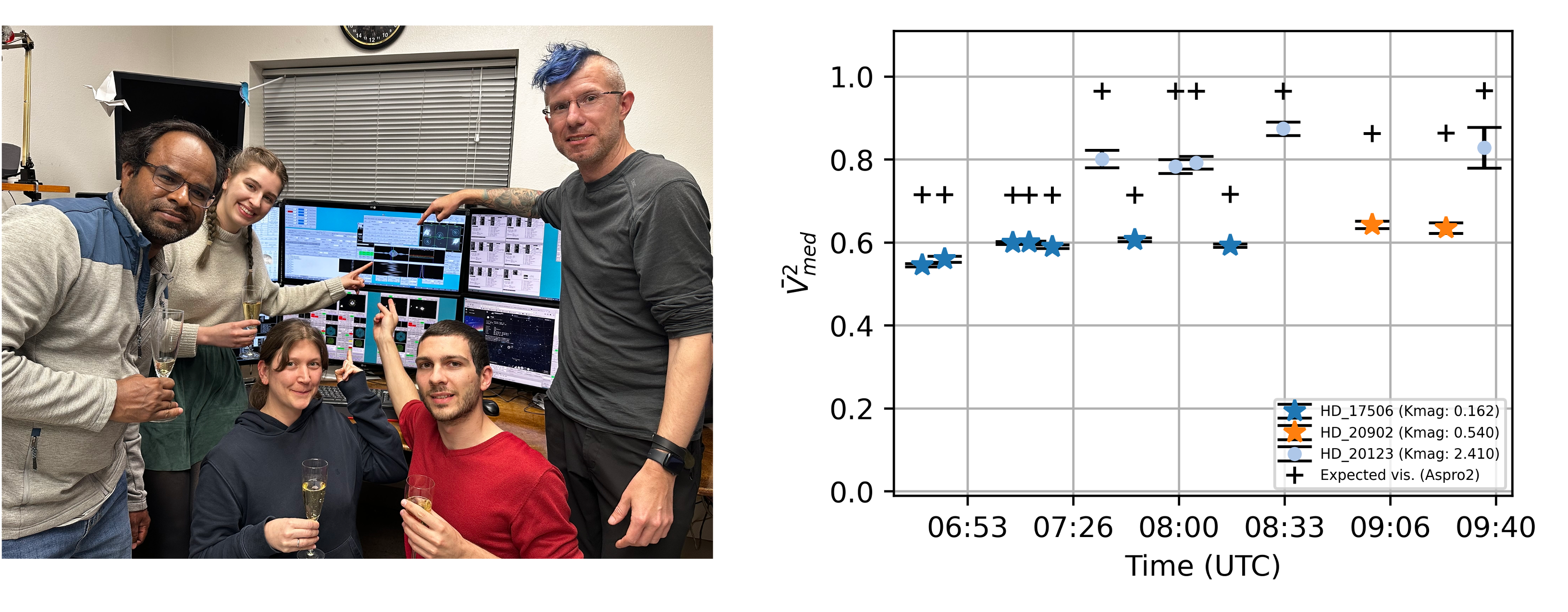}
    \vspace{0.25cm}
    \caption{Left: Picture celebrating the first fringes with CHARIOT at CHARA with some of the team members (here left to right: Narsi Anagu, Alyssa Mayer, Aline Dinkelaker, K\'evin Barjot, Nic Scott). Right: a three-hour observation sequence showing the stable difference between the theoretical square visibilities on spatially resolved stars (+ sign) and the observed visibilities (symbols with error bars). The difference corresponds to the transfer function.}
    \label{fig:Fig5}
\end{figure}

At CHARA, several engineering runs with access to the night sky were conducted over the last years to test and improve the performances of CHARIOT. After a first beam combiner was installed and first fringes observed in 2024, a second fibered-one replaced the previous one after a defect was found in the fiber gluing. During these runs, a large number of sources were observed in order to quantify the stability of the transfer function. Sources with magnitudes as faint as K=5 were repeatedly observed. An example of the typical observations conducted at CHARA is shown in Figure~\ref{fig:Fig5}. 

\subsection{Next steps at CHARA}

\begin{figure}[H]
    \centering
    \includegraphics[width=0.95\linewidth]{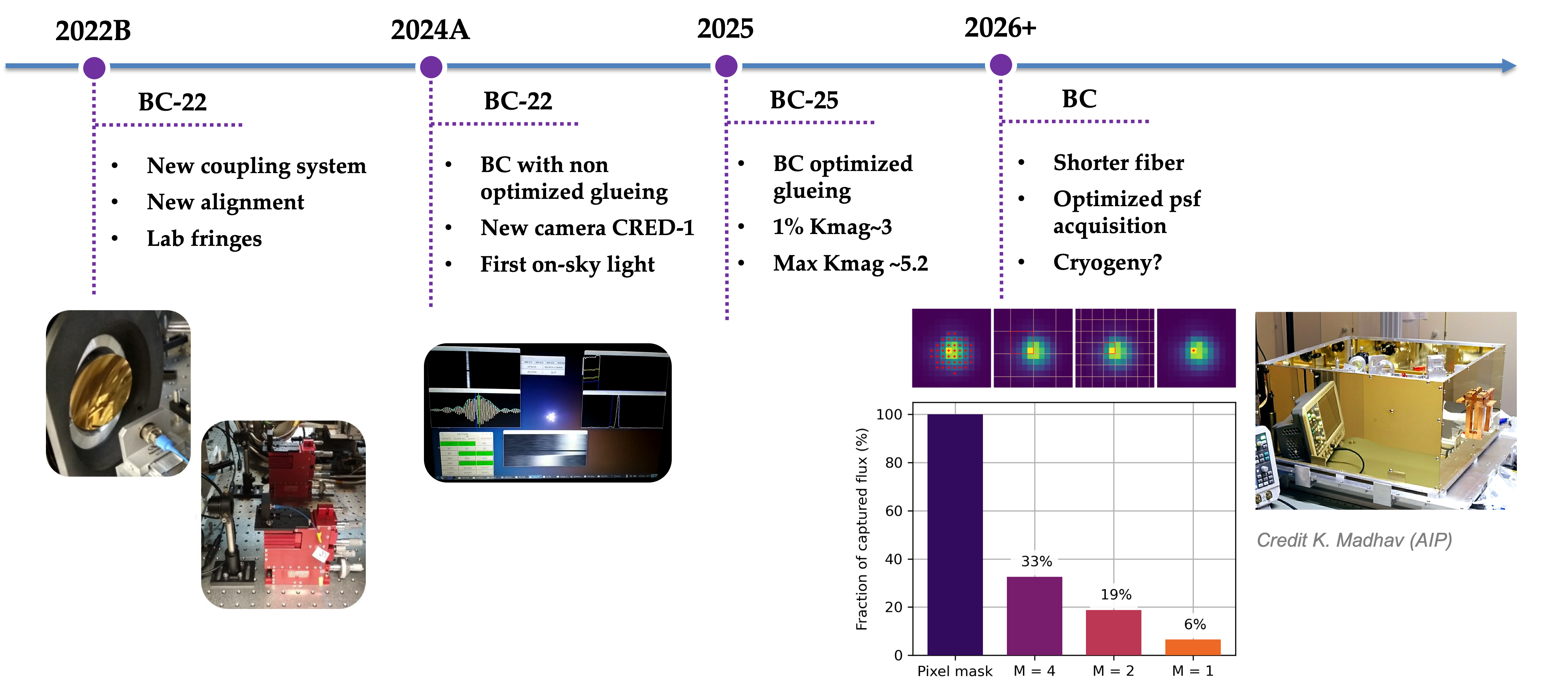}
    \vspace{0.25cm}
    \caption{Temporal visualization of the development of CHARIOT. Among the future tasks, a better readout scheme of the four output signals will allow to improve the SNR required for the derivation of the squared visibilities. Indeed, currently only a small fraction of the PSF flux is considered in the data reduction process.}
    \label{fig:Fig6}
\end{figure}

Room for improvement has already been identified in different areas of the CHARIOT instruments. Some are relatively straightforward to implement, whereas others concern the future upgrade of this new platform (extension to four telescope beam? Nulling inteferometry?). In Figure~\ref{fig:Fig6}, we illustrate a timeline showing some of the next steps. Furthermore, the reported years give a good idea of the overall effort to reach this point. It should be underlined that the biggest bottleneck in a rapid upgrade of CHARIOT is the access to the instrument by the team since only one to two missions per year could be organized. It is foreseeable that, with the implication of a larger team, the necessary critical mass for a fast turnaround could be reached. 

\section{Integral Field Unit based on Multicore fibers}

The MCIFU is a multicore integral field unit (IFU) based on novel fibre technology and is being developed as a pathfinder for future PCS technologies \cite{haffert2020, harris2020}\,. The instrument is planned for on-sky testing behind MagAO-X in 2027. A key objective is the development of a fibre link that combines a high core count ($>100$ spaxels), low inter-core crosstalk ($<10^{-3}$ between adjacent spaxels), and high throughput ($>50\%$ transmission).

\begin{figure}
    \centering
    \begin{tabular}{c c}
       \includegraphics[width=0.4\linewidth]{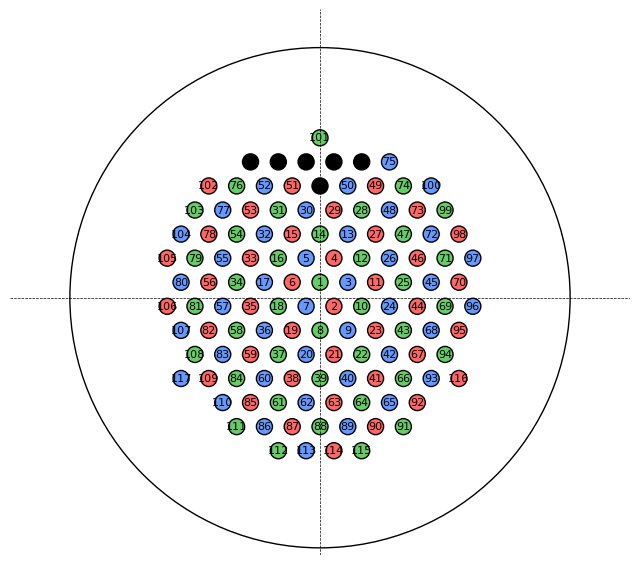}  & \includegraphics[width=0.4\linewidth]{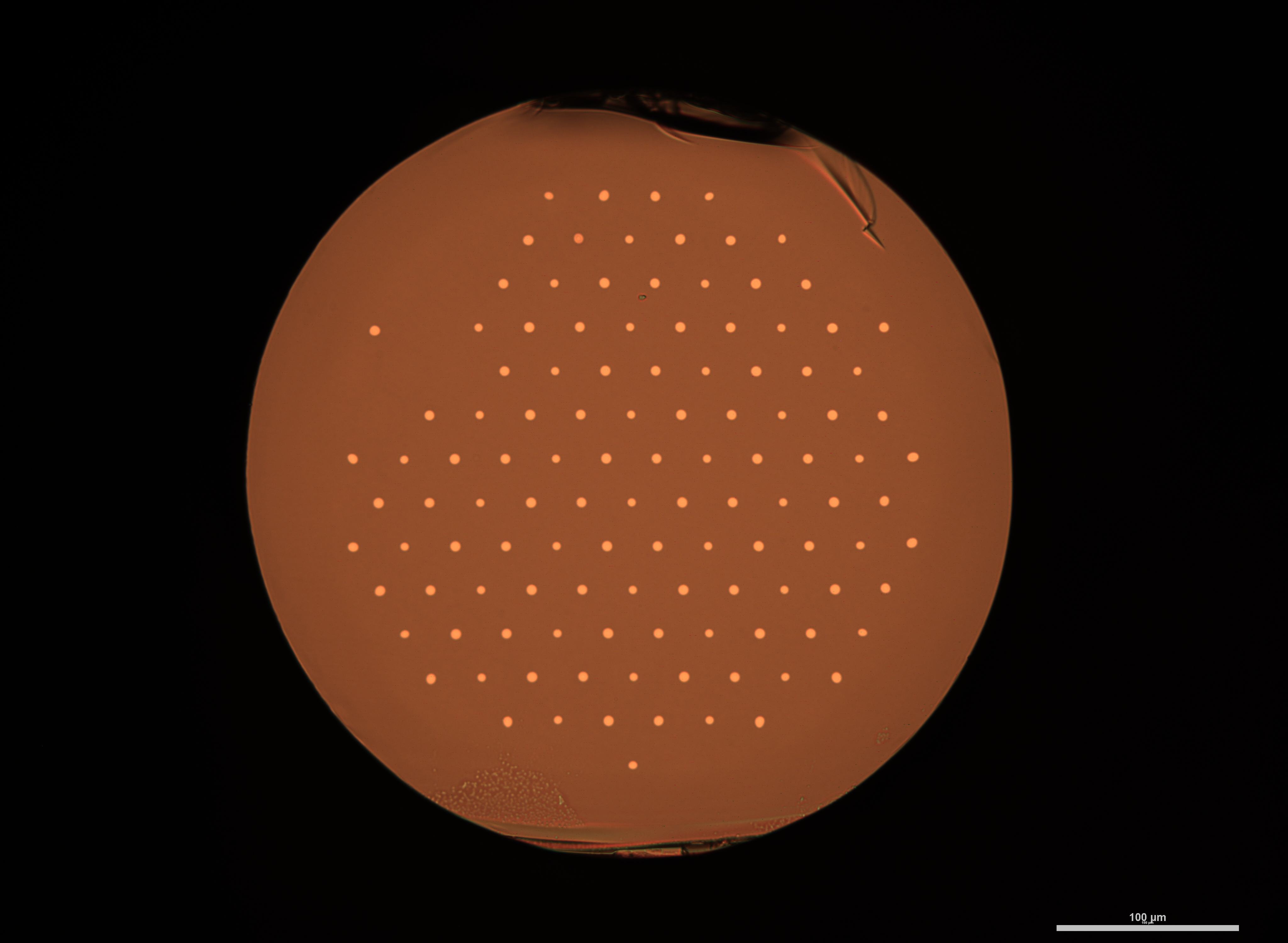} \\
    \end{tabular}    
        \vspace{0.25cm}
    \caption{The 117-core multicore fiber developed for the MCIFU project. Left) The theoretical design of the test fiber where the different colour cores represent the differing core sizes. Here black to note that no core is placed in this location and cladding glass is inserted to ensure proper fiber packing. The identical design was drawn to two different fiber diameters, resulting in different single mode wavelength ranges. Right) The drawn test fiber. Here you can clearly see the floating core and missing cores from the design. The differing size cores are also visible, along with imperfections due to the drawing procedure.}
    \label{fig:MCIFU_fibre}
\end{figure}

To meet these requirements, we have developed a heterogeneous multicore fibre, optimised for operation in the infrared (see Figure \ref{fig:MCIFU_fibre} and Mahdizadeh et al, proceedings of conference 14544). The use of heterogeneous cores suppresses crosstalk between neighbouring channels, allowing the cores to be packed more closely together while maintaining optical performance \cite{Koshiba2024}\,. Laboratory measurements indicate that crosstalk levels below $10^{-3}$ have been achieved (see Mahdizadeh et al). A ULI-fabricated photonic reformatter is currently being manufactured, and full system integration is planned for later this year.

\section{Small scale dispersers for high-level integration}

We investigated development of ultra-compact spectrographs fabricated using two-photon polymerisation (2PP) and printed directly onto the tip of a single-mode optical fibre \cite{ronson2026}\,. Each device is less than 1 mm in size and integrates all key optical functions, including beam expansion, collimation, dispersion, and stray-light suppression, into a monolithic structure. The concept aims to enable a new class of lightweight, alignment-free spectroscopic instruments that could be deployed in astronomy, quantum technologies, space payloads, and distributed sensing applications.

\begin{figure}
    \centering
    \begin{tabular}{c c}
       \includegraphics[width=0.3\linewidth]{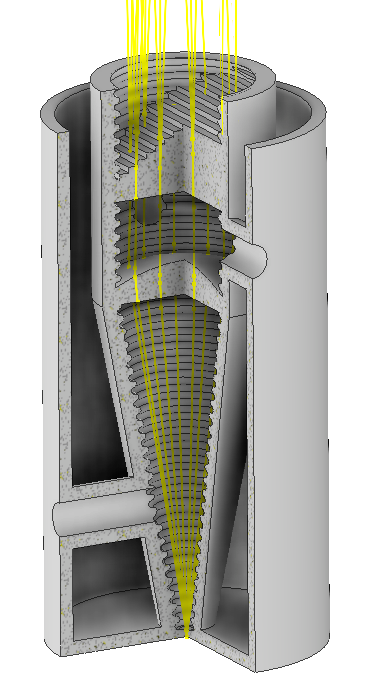}  & \includegraphics[width=0.3\linewidth]{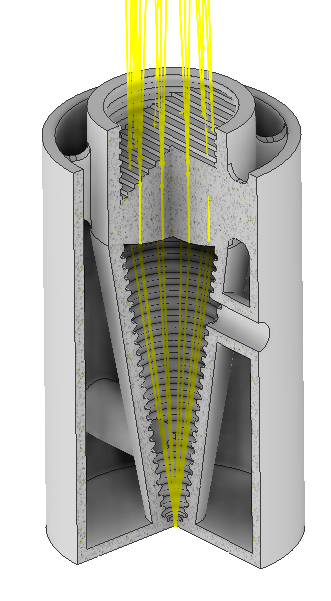} \\
    \end{tabular}    
            \vspace{0.25cm}
    \caption{CAD models of the two TPP spectrographs. Left) The multielement design has a height of 815 $\mu$m and outer diameter 200 $\mu$m, with a single element focussing and dispersing the light. Right) The single element design which has a height of 794 $\mu$m tall and an outer diameter of 240 $\mu$m. Reproduced from Ronson et al 2026 \cite{ronson2026}.}
    \label{fig:spectrograph}
\end{figure}

Two spectrograph architectures were investigated: a multi-element design, consisting of separate beam-expansion, collimation, and grism elements, and a single-element design that combines the collimator and dispersive optics into a single printed component (see Figure~\ref{fig:spectrograph}). Both designs were optimised for near-infrared operation between approximately 1.0 and 1.7 $\mu$m and targeted a modest spectral resolving power suitable for proof-of-concept demonstrations. The devices were manufactured using high-resolution 2PP printing and incorporated blackened cavities and roughened internal surfaces to minimise stray light.

Laboratory characterisation demonstrated that both spectrographs function successfully as fully integrated fibre-coupled instruments. Transmission measurements showed clear benefits from reducing the number of optical interfaces, with the single-element design achieving approximately $80\%$ throughput across most of the operating band, compared with $50-60\%$ throughput for the multi-element design. Spectral testing yielded resolving powers of approximately R $\approx$ 20-26, exceeding the original proof-of-concept requirement and confirming the feasibility of printed micro-spectrographs for near-infrared spectroscopy.

These results validate the concept of millimetre-scale spectrographs and highlight the advantages of additive manufacturing for highly integrated photonic instrumentation. Future work will focus on understanding the gap between measured and theoretical resolving power, improving stray-light performance, testing multiple device variants, and exploring reflective spectrograph architectures. The demonstrated combination of compact size, high throughput, and direct fibre integration suggests significant potential for future astronomical and photonic instrumentation where size, mass, and scalability are critical constraints

\section{Printed Pyramids}

Pyramid wavefront sensors are among the most sensitive wavefront sensing technologies available for high-contrast astronomy, but conventional glass pyramids are difficult and expensive to manufacture because they require extremely sharp edges, high surface quality, and precise geometries. We aim to demonstrate that 3D optical printing can overcome many of these limitations, enabling compact, low-cost pyramids with geometries that are difficult or impossible to realize using traditional polishing techniques (see Magniez et al. 2026, proceedings of conference 14544). Four prototype inverted pyramids with diameters of 0.5 mm and 1 mm and wedge angles of 0.5° and 2° were fabricated and tested.

\begin{figure}
    \centering
    \includegraphics[width=0.3\linewidth]{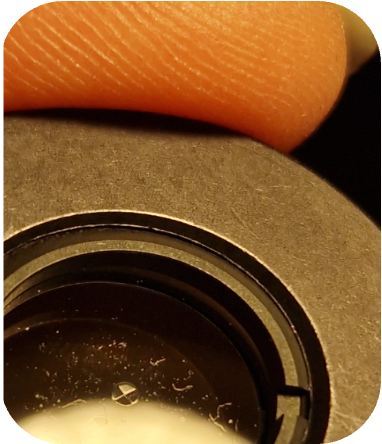} \\
            \vspace{0.25cm}
    \caption{ The fabricated pyramid sensor mounted in a 1 inch optic mount. For further details see Magniez et al. 2026, proceedings of conference 14544}
    \label{fig:pyramid}
\end{figure}

A detailed metrology campaign showed that the printed pyramids meet or exceed key manufacturing requirements. Measurements from both the manufacturer and independent Zygo surface-profiling tests demonstrated wedge angles accurate to specification, surface roughness between approximately 5 and 21 nm RMS, and edge transition regions approaching the sub-micron scale. Such performance is comparable to, and in some respects better than, traditionally fabricated pyramid optics. The printed pyramids also benefit from being extremely thin, reducing chromatic effects that often complicate glass-based pyramid wavefront sensors.

Because the current printed pyramids are physically small, they naturally introduce a form of spatial filtering. Numerical simulations showed that low spatial frequency aberrations are measured by the conventional four-pupil pyramid signal, while higher spatial frequencies bypass the pyramid and form an additional central pupil image. This creates a novel sensing architecture that can simultaneously probe different spatial frequency regimes and may provide improved sensitivity compared with conventional modulated pyramid sensors. The models suggest that such spatially filtered pyramids could retain high sensitivity while reducing some of the complexity associated with modulation.

Initial laboratory testing on Durham University's POPY adaptive optics testbed successfully demonstrated open-loop wavefront sensing with the printed pyramids, confirming the predicted appearance of both conventional and central pupil signals. Future work includes fabrication of larger anti-reflection-coated pyramids, closed-loop adaptive optics operation, on-sky demonstrations with the REVOLT test facility in Canada, and eventual integration into the planned Keck Observatory infrared pyramid wavefront sensor upgrade. More broadly, the work highlights how additive manufacturing can enable entirely new classes of wavefront sensor designs, including flattened pyramids, double-pyramid architectures, and other advanced concepts that are difficult to realize with conventional optics.

\section{TPP Long-term performance validation}

We performed a six-year study of the long-term performance of a two-photon polymerized microlens ring tip-tilt sensor (MLR-TT) installed at the Large Binocular Telescope (LBT) (see Harris et al. 2026, proceedings of conference 14544). The sensor was developed to improve single-mode fibre coupling by measuring residual tip-tilt errors that are not corrected by the telescope's adaptive optics system. A microlens ring printed directly onto a fibre assembly redirects a small fraction of the incoming light into surrounding multimode sensing fibres while maintaining efficient coupling into the central science fibre. The primary aim of the study was to assess whether two-photon polymerization (TPP) fabricated micro-optics can survive and maintain performance under realistic observatory conditions over timescales comparable to those required for astronomical instrumentation.

Repeated measurements obtained between 2019 and 2026 demonstrate that the sensor remains operational and exhibits a highly consistent response despite prolonged exposure to temperature variations, humidity, vibration, contamination, and telescope motion. Circular beam-scanning tests showed nearly identical behavior across multiple epochs, with minor differences attributed to instrument alignment changes rather than degradation of the printed optic itself. These results represent one of the longest in-situ telescope deployments of a two-photon-polymerized optical component and provide strong evidence that 3D-printed micro-optics can deliver the robustness and stability required for long-term astronomical use. The study provides an important validation of additive manufacturing as a viable technology for future astronomical instrumentation.


\acknowledgments 

This research has received funding from the Deutsche Forschungsgemeinschaft (DFG), grant number 506421303 ("NAIR-APREXIS"), the Bundesministerium für Bildung und Forschung (BMBF), grant number 03Z22AI1 ("Strategic Investment"), the H2020 Future and Emerging Technologies (820365-PHoG), the Science and Technology Facilities Council (ST/V000403/1), the Engineering and Physical Sciences Research Council (EP/X03299X/1), and the European Union's Horizon 2020 research and innovation programme under Grant Agreement 101004719 (ORP). 
The CHARA Array is supported by the National Science Foundation under Grant No. AST-2034336 and AST-2407956. Institutional support has been provided from the GSU College of Arts and Sciences, Office of the Provost, and Office of the Vice President for Research and Economic Development.

\bibliography{report} 
\bibliographystyle{spiebib} 

\end{document}